\documentclass[fleqn,usenatbib,useAMS]{mnras}

\usepackage{graphicx}	% Including figure files
\usepackage{amsmath}	% Advanced maths commands
\usepackage{amssymb}	% Extra maths symbols
\usepackage{multicol}        % Multi-column entries in tables
\usepackage{bm}		% Bold maths symbols, including upright Greek
\usepackage{pdflscape}	% Landscape pages

\usepackage[T1]{fontenc}
\usepackage{ae,aecompl}

\usepackage{newtxtext,newtxmath}

\title{Light Curve Analysis of the AP Dor Binary System using Ground-Based and TESS Observations}

\author[Poro et al.]{A. Poro$^{1,}$\thanks{Corresponding author: poroatila@gmail.com}%
, E. Fernández-Lajús$^{2,3}$,
M. Madani$^{4}$,
G. Sabbaghian$^{4}$,
F. Nasrollahzadeh$^{4}$,
F. Jahediparizi$^{5}$
\\
\\
$^{1}$Astronomy Department of the Raderon AI Lab., BC., Burnaby, Canada\\
$^{2}$Instituto de Astrofísica de La Plata (CCT La Plata-CONICET-UNLP), La Plata, Argentina\\
$^{3}$Facultad de Ciencias Astronómicas y Geofísicas, Universidad Nacional de La Plata, Paseo del Bosque, B1900FWA, La Plata, Argentina\\
$^{4}$BSN Project, Transiting Exoplanets Department, Tehran, Iran\\
$^{5}$Department of Physics, Shahid Bahonar University of Kerman, Kerman, Iran}

\date{Accepted ---. Received ---; in original form ---}

\pubyear{Draft Version}

\begin{document}
\label{firstpage}
\pagerange{\pageref{firstpage}--\pageref{lastpage}}
\maketitle

\begin{abstract}
 The short-period AP Dor eclipsing binary's first in-depth and multiband photometric solutions are presented. We made use of our eight nights of ground-based at a southern hemisphere observatory, and twelve sectors of TESS observations. We extracted eight and 1322 minima from our observations and TESS, respectively. We suggested a new linear ephemeris based on the trend of orbital period variations using the Markov chain Monte Carlo (MCMC) approach. The PHysics Of Eclipsing BinariEs (PHOEBE) Python code and the MCMC approach were used for the light curve analysis. This system did not require a starspot for the light curve solutions. We calculated the absolute parameters of the system using $Gaia$ DR3 parallax method. The orbital angular momentum ($J_0$) of the AP Dor indicates that this system is located in a region of contact binaries. According to our results, this system is an overcontact binary system with a mass ratio of 0.584, a fillout factor of 48\%, and an inclination of $53^{\circ}$. The positions of AP Dor stars on the Hertzsprung-Russell (HR) diagram are represented.
\end{abstract}


%%%%%%%%%%%%%%%%%%%%%%%%%%%%%%%%%%%%%%%%%%%%%%%%%%

\section{Introduction}
Eclipsing binaries are a significant astrophysical tool for investigating star formation, stellar structure, and the physical properties of stars and their evolution.

Both stars in a binary system known as an overcontact binary have exceeded their Roche lobes. Due to the tidally distorted forms of the stars, the light curve of an overcontact system is continuously changeable and is typically categorised as being of the W UMa type. Mass transfer through Lagrange points is likely to happen in such a system. Other features of them are that the temperatures of the components are roughly equal because they share a similar envelope with the same entropy (\citealt{2006MNRAS.368.1311P}).

W UMa stars also known as the low-mass eclipsing binaries consisting of ellipsoidal components with orbital periods less than $1^{day}$, usually $P<0.7^{day}$ (\citealt{2022MNRAS.510.5315P}).
\\
\\
The AP Dor (HIP 023793) binary system has an apparent magnitude of 9.37\footnote{\url{http://simbad.cds.unistra.fr/simbad}} and is located in the southern hemisphere with coordinates R.A.: $05^h$ $06^m$ $45.09188^s$ and Dec: $-59^\circ$ $03'$ $03.45465"$ (J2000).

This system is introduced as an EW\footnote{W Ursae Majoris-type eclipsing variables} type in the VSX\footnote{\url{https://www.aavso.org/vsx/}} database with an orbital period of 0.427187 days, but its orbital period is unknown in the ASAS-SN\footnote{\url{https://asas-sn.osu.edu/variables}} catalog.

For the first time, this system is classified as a W UMa-type system or possibly a RR Lyrae star in the $HIPPARCOS$ catalog. In a subsequent study, \cite{1980IBVS.1772....1E} introduced it as a contact system. Then, by the \cite{2004AA...416.1097S} study, three main geometric parameters ($q$, $f$, and $i$) have been estimated for 64 $HIPPARCOS$ catalog contact systems, including AP Dor.
\\
\\
The structure of the paper is as follows: Section 2 provides details on photometric observations and a data reduction method. Section 3 presents the minima and the new ephemeris of the AP Dor system. The photometric light curve solutions for the system are discussed in Section 4. Section 5 provides a description of the method used to determine absolute parameters. At the end, Section 6 includes the summary and conclusion.

%%%%%%%%%%%%%%%%%%%%%%%%%%%%%%%%%%%%%%%%%%%%%%%%%%

\section{Observation and Data Reduction}
The photometric observations of AP Dor were carried out on October 24-31, 2017, and a total of 2897 images were taken in eight nights. These observations were made using the 0.60m "Helen Sawyer Hogg" (HSH) telescope at the Complejo Astronomico El Leoncito (CASLEO) Observatory, Argentina ($69^\circ 18’$ W, $31^\circ 48’$ S, 2552m above sea level). A CCD SBIG STL1001E and $BVRI$ standard filters were employed. The average temperature of CCD during the observation nights was $-30^\circ$C. Each of the frames was $1\times1$ binned with averaged 50s exposure time for the $B$ filter, 45s for the $V$ filter, 40s for the $R$ filter, and 30s for the $I$ filter.

UCAC4 156-005107 was selected as a comparison star and TYC 8517-653-1 was chosen as a check star. The comparison star was found at R.A. $05^h$ $07^m$ $13.49^s$, Dec. {$-58^\circ$ $59'$ $58.59"$} (J2000) with a $V=12.203(30)$ magnitude, while the check star was located at R.A. $05^h$ $06^m$ $35.85^s$, Dec. {$-58^\circ$ $57'$ $53.24"$} (J2000) with a $V=11.57(12)$ magnitude, according to the Simbad astronomical database.

The APPHOT photometry package of the Image Reduction and Analysis Facility\footnote{\url{http://iraf.noao.edui}} (IRAF) was used for CCD reduction and aperture photometry.

The Transiting Exoplanet Survey Satellite (TESS) mission observed the AP Dor system in sectors 1, 4, 6, 8, 10, 13, 27, 30, 31, 32, 34, and 39. TESS data is available at the Mikulski Space Telescope Archive (MAST)\footnote{\url{http://archive.stsci.edu/tess/all products.html}}. The LightKurve code\footnote{\url{https://docs.lightkurve.org}} was used to extract TESS style curves from the MAST, which had been detrended using the TESS Science Processing Operations Center (SPOC) pipeline (\citealt{2016SPIE.9913E..3EJ}).

\section{Orbital Period Variations}
We used a Python code using a Gaussian function and the MCMC method to extract the new mid-eclipse times and uncertainty. The code is implemented in Python using the PyMC3 package (\citealt{2016ascl.soft10016S}). Therefore, we extracted eight mid-eclipse times, including four primary and four secondary minima, from our observations, with two times recorded for each $BVRI$ filters (Table \ref{tab1}). In addition, we extracted a total of 1322 minima from different sectors of TESS observations (Appendix Table \ref{A1}). We found two minima from the \cite{2017OEJV..179....1J} study, and we added them as literature. Barycentric Julian Date in Barycentric Dynamical Time ($BJD_{TDB}$) was used to express all minimum times. We used OSU Online Astronomy Utilities\footnote{\url{https://astroutils.astronomy.osu.edu/time/hjd2bjd.html}} to convert the literature minimum times to $BJD_{TDB}$.

There are two different orbital periods for this system in the catalogs: in the AAVSO catalog, the value is $0.427187^d$, and in the ASAS3 catalog, the value is $0.213593^d$. Based on a Fourier analysis of our data, we conclude that the AAVSO catalog value is more valid. Therefore, we used the orbital period of $0.427187^d$ along with the minimum time from our observation as a reference ephemeris.

\begin{figure*}
\begin{center}
\includegraphics[scale=0.40]{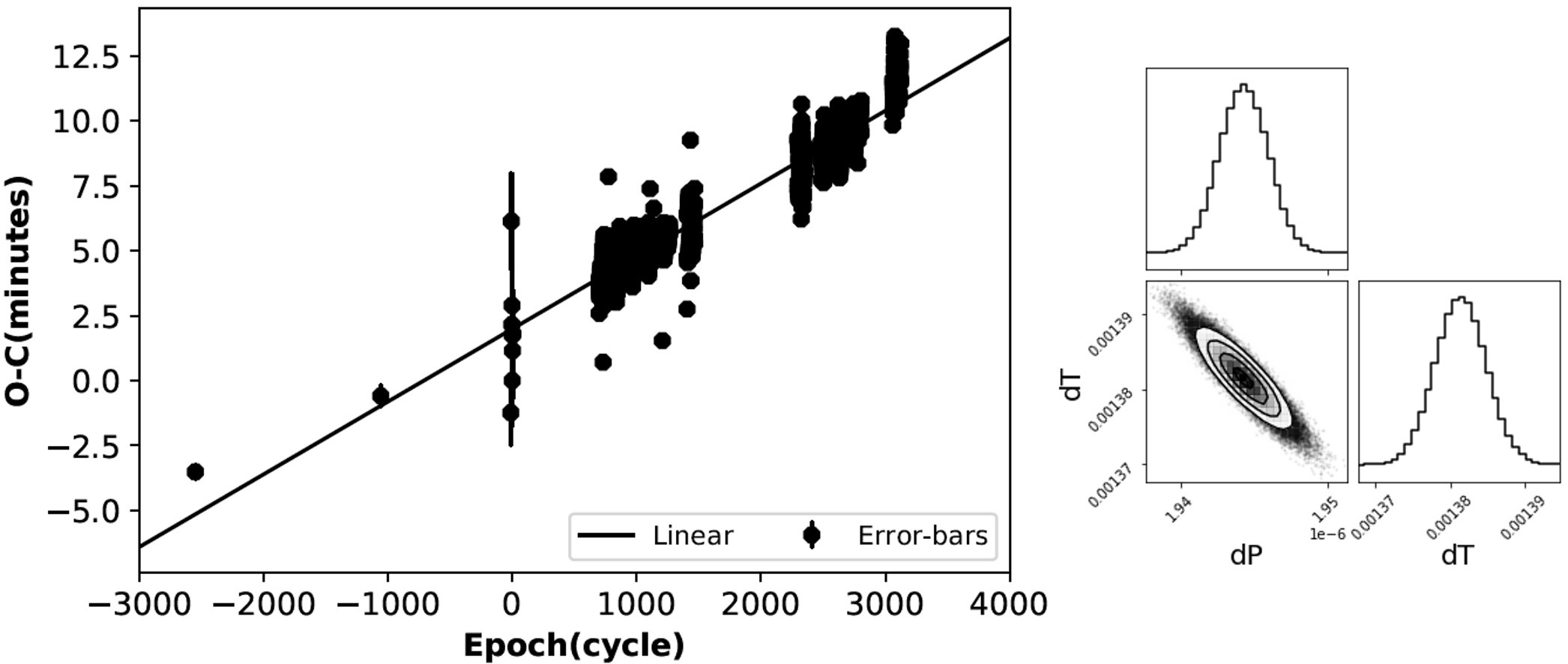}
    \caption{The O-C diagram of the AP Dor binary system is on the left, and the corner plot obtained from MCMC is on the right.}
\label{Fig1}
\end{center}
\end{figure*}

The O-C variations are the observed mid-eclipse times (O) from their calculated values (C) based on a reference ephemeris. Typically, a trend in these variations is the result of several separate effects working together. Figure \ref{Fig1} shows the O-C diagram of the AP Dor system. According to the visible trend for the O-C diagram, only a linear fit can be considered. We calculate a new ephemeris based on the MCMC method using the Emcee package in Python (\citealt{2013PASP..125..306F}). We applied 20 walkers and 20,000 iterations for each walker, with a 1000 burn-in period in the MCMC sampling. Due to the linearity of the fit, the values of orbital period and minimum time were considered priors from the reference ephemeris.

\begin{table}
\caption{Times of minima based on the ground-based $BVRI$ observations.}
\centering
\begin{center}
\footnotesize
\begin{tabular}{c c c c c}
 \hline
 \hline
 Min.($BJD_{TDB}$)	& Error	& Filter & Epoch & O-C\\
\hline
2458050.708890	&	0.000873	&	$V$	&	-11.5	&	-0.00086	\\
2458052.636356	&	0.001302	&	$V$	&	-7	&	0.00426	\\
2458053.701563	&	0.000783	&	$I$	&	-4.5	&	0.00150	\\
2458054.770044	&	0.001515	&	$I$	&	-2	&	0.00201	\\
2458055.622405	&	0.001227	&	$R$	&	0	&	0.00000	\\
2458055.837207	&	0.001296	&	$R$	&	0.5	&	0.00121	\\
2458056.691174	&	0.001276	&	$B$	&	2.5	&	0.00080	\\
2458057.759589	&	0.001168	&	$B$	&	5	&	0.00125	\\
\hline
\hline
\end{tabular}
\end{center}
\label{tab1}
\end{table}

The following light elements were assigned to a new revised linear ephemeris for the minima obtained from this study, TESS, and the literature:

\begin{equation}
\label{eq1}\begin{aligned}
BJD_{TDB}(Min.I)=2458055.623786(3)+0.427188944(2)\times E
\end{aligned}
\end{equation}

where $E$ is the integer number of orbital cycles after the reference epoch. The upper and lower limits of uncertainties for the elements in MCMC were equal.

\section{Light Curve Analysis}
Light curve analysis of the AP Dor binary system was carried out using the PHOEBE 2.4.9 version and the MCMC approach (\citealt{2005ApJ...628..426P}, \citealt{2020ApJS..250...34C}, \citealt{2022PASP..134f4201P}). We selected contact mode for the light curve solutions based on how the light curve appeared and the system's short orbital period.

The gravity-darkening coefficients and the bolometric albedo were assumed to be $g_1=g_2=0.32$ (\citealt{1967ZA.....65...89L}) and $A_1=A_2=0.5$ (\citealt{1969AcA....19..245R}), respectively. The limb-darkening coefficients were used as free parameters in PHOEBE, and the \cite{2004A&A...419..725C} method was used to model the stellar atmosphere. Regarding the primary star's temperature input, we tried three methods and compared the results. However, based on our observational data, we set the value obtained from $B-V$ as the temperature of the primary star. So, after the required calibrations (\citealt{2000AA...357..367H}) we calculated $(B-V)_{AP Dor}=0.428\pm0.013$ and the effective temperature of the primary component, $T_1$ was assumed as $6517\pm121$ (\citealt{2020MNRAS.496.3887E}). We calculated $T_1$ from the relationship between the primary star temperature and the orbital period of the system from the study of \cite{2022MNRAS.510.5315P} to be $6396\pm92$. Also, the temperature of the system is determined by $Gaia$ DR2 to be $6530_{\rm-159}^{+129}$.

One of the most important input parameters to the PHOEBE code is the mass ratio. We ran a $q$-search with PHOEBE and then used the code's optimization tool to improve the results. As a result, preliminary analyses were improved using the MCMC method, and the uncertainty estimates were obtained (Table \ref{tab2}). In the MCMC approach based on the Emcee package, we applied 96 walkers and 800 iterations to each walker in the MCMC approach. It should be noted that the light curve solution for this system did not require adding a star spot.

\begin{table}
\caption{Light curve solutions of AP Dor.}
\centering
\begin{center}
\footnotesize
\begin{tabular}{c c c}
 \hline
 \hline
Parameter && Result\\
\hline
$T_{1}$ (K) && $6585_{\rm-(39)}^{+(51)}$\\
\\
$T_{2}$ (K) && $6302_{\rm-(29)}^{+(39)}$\\
\\
$q=M_2/M_1$ && $0.584_{\rm-(13)}^{+(8)}$\\
\\
$\Omega_1=\Omega_2$ && $2.866(171)$\\
\\
$i^{\circ}$ &&	$53.00_{\rm-(13)}^{+(12)}$\\
\\
$f$ && $0.4875_{\rm-(96)}^{+(127)}$\\
\\
$l_1/l_{tot}$ && $0.644_{\rm-(2)}^{+(3)}$\\
\\
$l_2/l_{tot}$ && $0.356_{\rm-(2)}^{+(3)}$\\
\\
$r_{1(mean)}$ && $0.467(50)$\\
\\
$r_{2(mean)}$ && $0.375(41)$\\
\\
Phase shift && $0.006(2)$\\
\hline
\hline
\end{tabular}
\end{center}
\label{tab2}
\end{table}

\begin{figure*}
\begin{center}
\includegraphics[scale=0.76]{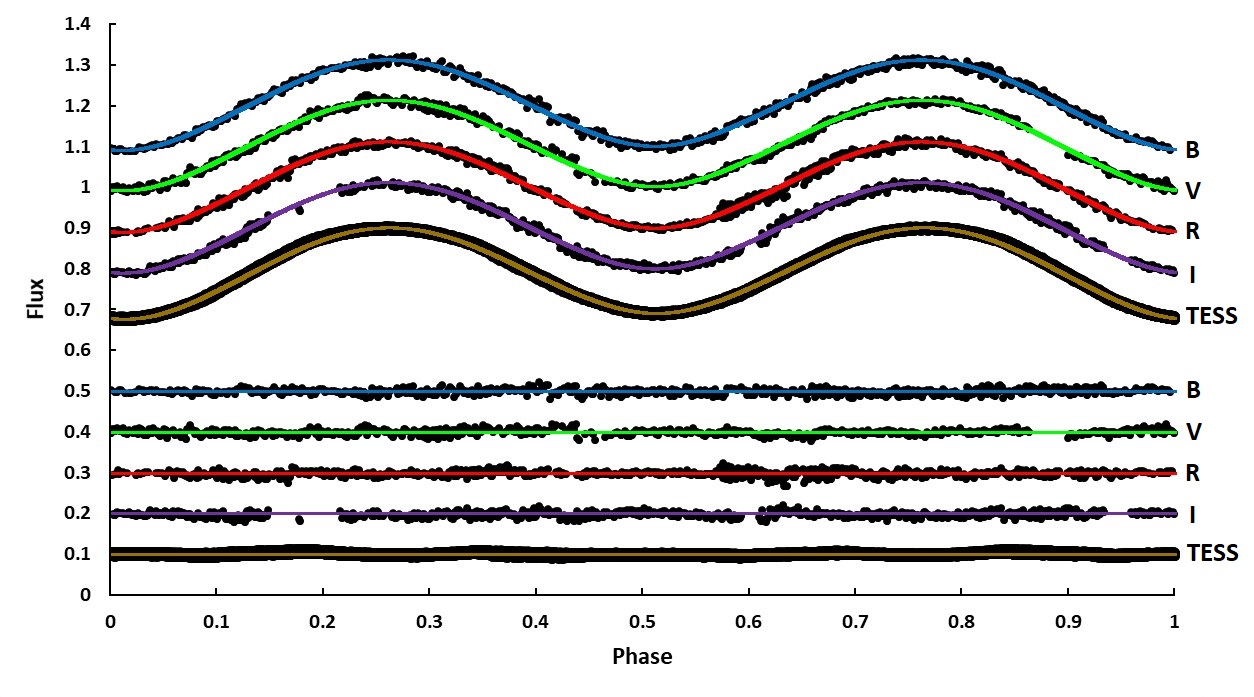}
    \caption{The observed light curves of AP Dor (black dots), and synthetic light curves obtained from light curve solutions in the $BVRI$ filters and TESS (top to bottom respectively); with respect to orbital phase, shifted arbitrarily in the relative flux.}
\label{Fig2}
\end{center}
\end{figure*}

\begin{figure*}
\begin{center}
\includegraphics[scale=0.30]{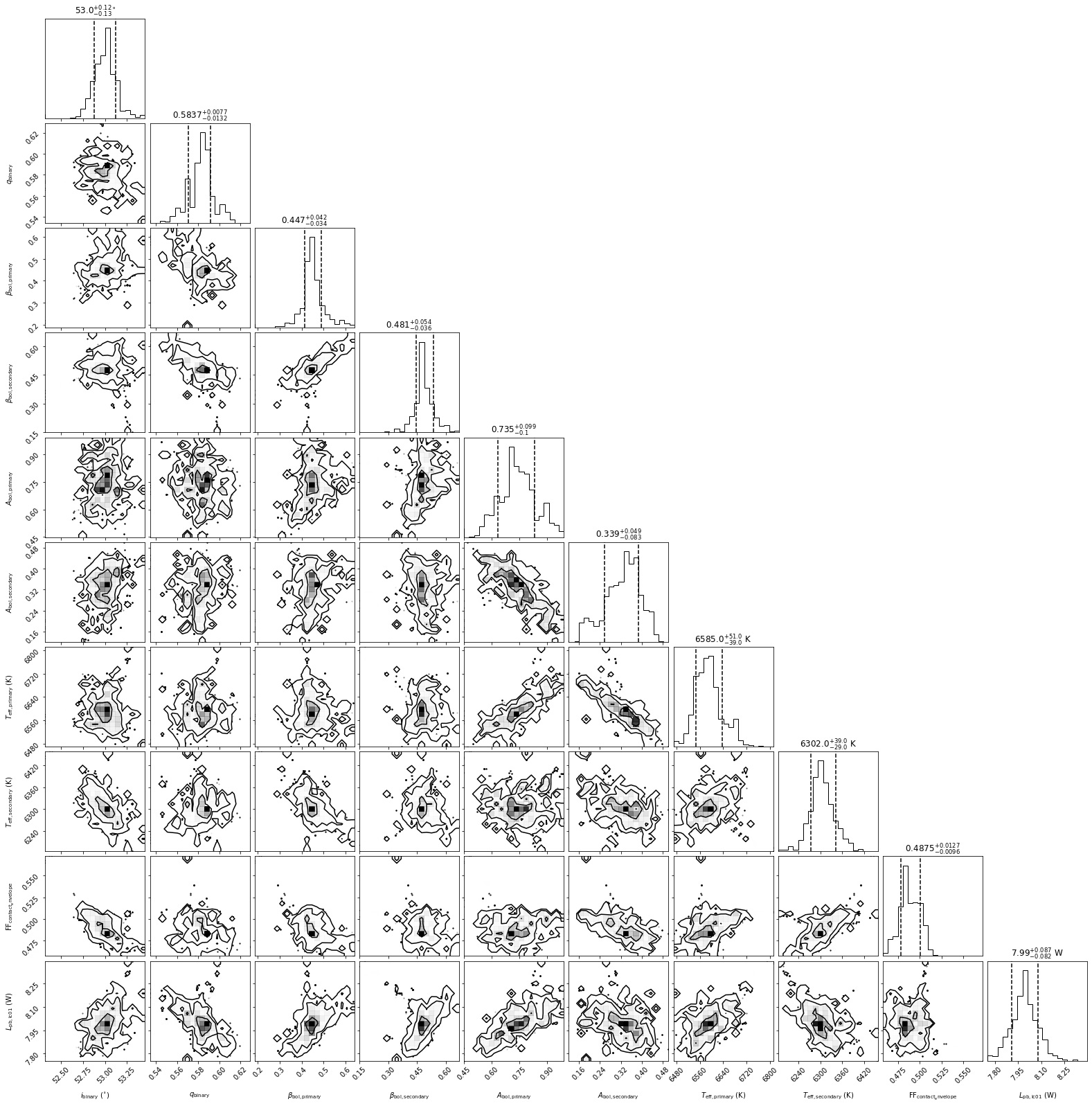}
    \caption{The corner plots of the AP Dor system was determined by MCMC modeling.}
\label{Fig3}
\end{center}
\end{figure*}

\begin{figure*}
\begin{center}
\includegraphics[scale=0.32]{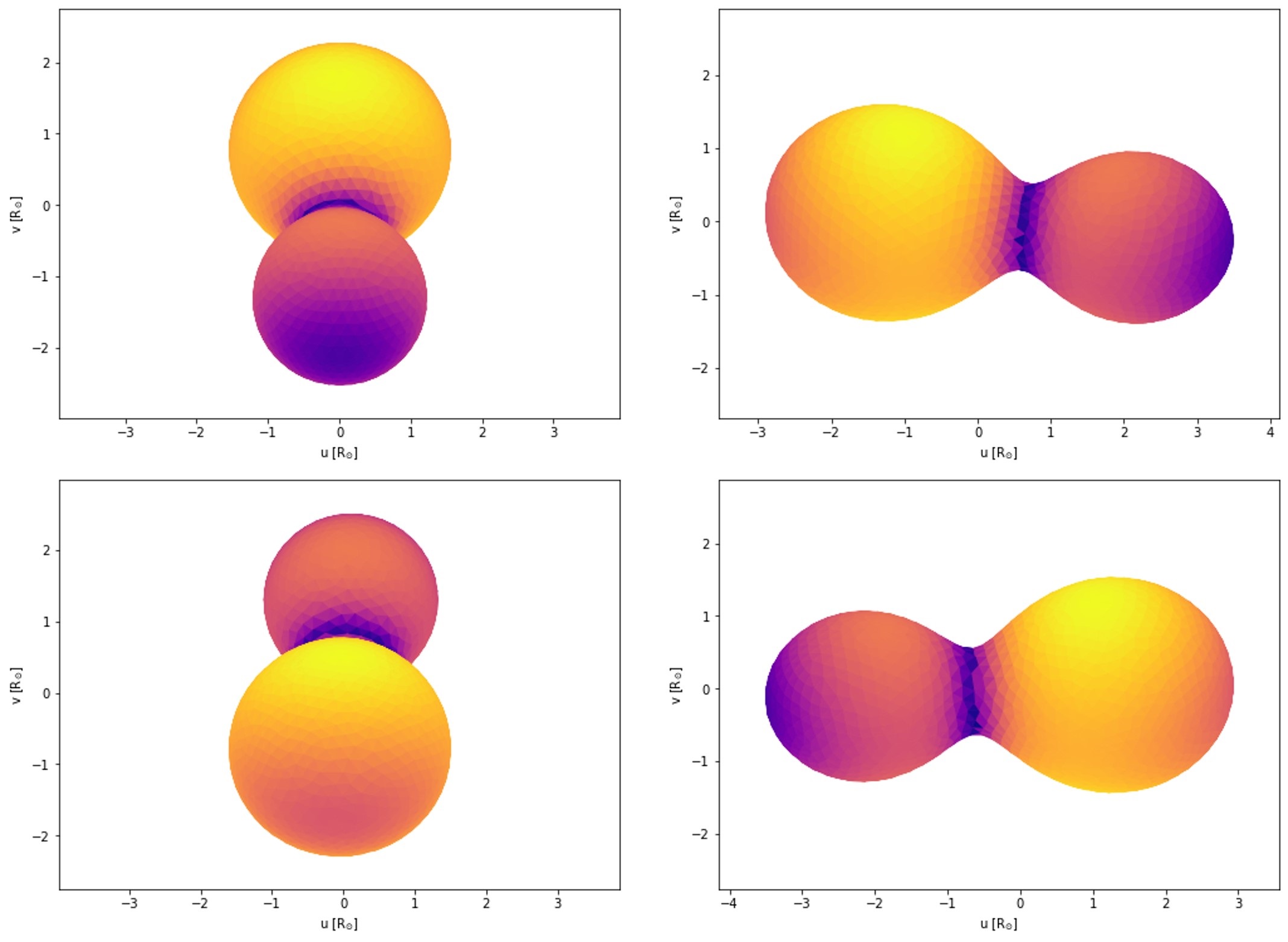}
    \caption{3D view of the AP Dor system stars.}
\label{Fig4}
\end{center}
\end{figure*}

The observed and theoretical light curves are shown in Figure \ref{Fig2}. The corner plot that MCMC produced is displayed in Figure \ref{Fig3}. The geometrical structure and 3D view of the AP Dor binary system are provided in Figure \ref{Fig4}.

\section{Absolute Parameters}
When just photometric data are available, one of the possible ways for estimating absolute parameters is to use the parallax $Gaia$ DR3 method. The method of calculating the parameters is described in the study of \cite{2022MNRAS.510.5315P}, and the parameters $d$(pc), $A_v$, $V_{max}$(mag), $l_{1,2}/l_{tot}$, $BC_{1,2}$, $T_{1,2}$, $r_{mean1,2}$, and $P$(day) is needed for this estimation. Accordingly, $M_{v(system)}$, $M_{v1,2}$, $M_{bol1,2}$, $L_{1,2}$, $R_{1,2}$, $a_{1,2}$, $M_{1,2}$ calculated, respectively. The separation $a$ is the average value of a1 and a2 calculated for each component; $a_1$ and $a_2$ must be close to each other, otherwise, it is not possible to use this method to calculate absolute parameters. We utilized $V_{max}=9.34(4)$ from our observations, the extinction coefficient $A_v=0.035(1)$ from the \cite{2011ApJ...737..103S} study, the system's distance from $Gaia$ DR3 $d_{(pc)}=186.402(398)$ to accomplish the estimation of the absolute parameters. Also, each star's bolometric magnitude was calculated using $BC_1=0.074$ and $BC_2=0.057$ from \cite{2020MNRAS.496.3887E} study. The results of the $Gaia$ DR3 method for estimating the absolute parameters of the AP Dor system are given in Table \ref{tab3}.
In addition, $M_{bol1,2}$, $logg_{1,2}$, and $a(R_\odot)$ parameters have been calculated using the following well-known equations respectively:

\begin{equation}\label{eq2}
M_{bol}-M_{bol_{\odot}}=-2.5log(\frac{L}{L_{\odot}})
\end{equation}

\begin{equation}\label{eq3}
g=G_{\odot}(M/R^2)
\end{equation}

\begin{equation}\label{eq4}
\frac{P^2}{4\pi^2}=\frac{a^3}{G(M_1+M_2)}
\end{equation}

\begin{table}
\caption{Estimation of the AP Dor's absolute parameters.}
\centering
\begin{center}
\footnotesize
\begin{tabular}{c c c c c}
 \hline
 \hline
Parameter & & Primary & & Secondary\\
\hline
$M(M_\odot)$ && $1.278_{\rm-(75)}^{+(123)}$ && $0.746_{\rm-(60)}^{+(83)}$\\
$R(R_\odot)$ && $1.360_{\rm-(31)}^{+(31)}$ && $1.113_{\rm-(10)}^{+(40)}$\\
$L(L_\odot)$ && $3.119(85)$ && $1.752(43)$\\
$M_{bol}(mag)$ && $3.505(30)$ && $4.131(27)$\\
$log(g)(cgs)$ &&	$4.277_{\rm-(24)}^{+(39)}$ && $4.218_{\rm-(34)}^{+(45)}$\\
$a(R_\odot)$ && $3.019_{\rm-(37)}^{+(30)}$ &&\\
\hline
\hline
\end{tabular}
\end{center}
\label{tab3}
\end{table}

\section{Conclusion}
The AP Dor short-period binary system was observed during a period of eight nights at a southern hemisphere observatory using $BVRI$ standard filters. We extracted times of minima from our observations and TESS data and presented a new ephemeris for the system using the MCMC method. The O-C diagram displayed a linear and increasing trend.

\begin{figure*}
\begin{center}
\includegraphics[scale=0.33]{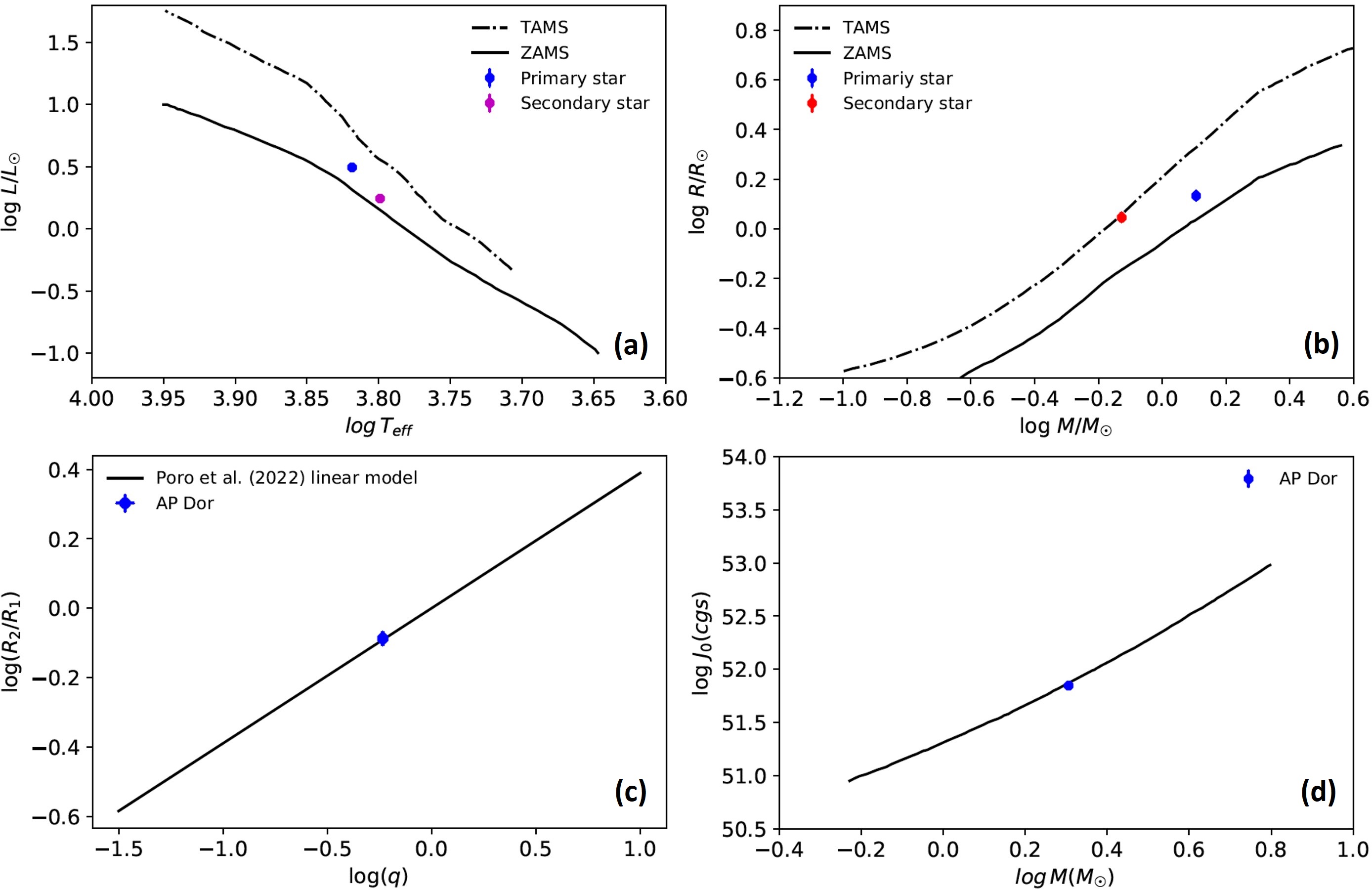}
    \caption{The position of AP Dor on the a) HR, b) $logR-logM$, c) $logR_{ratio}-logq$, d) $logJ_0-logM$ diagrams, respectively.}
\label{Fig5}
\end{center}
\end{figure*}

Utilizing PHOEBE Python code and the MCMC approach, the light curves of this system were analyzed. There is a 283 K temperature difference between the two components. These temperatures indicate that the primary and secondary components' spectral types are F5 and F7, respectively (\citealt{2000asqu.book.....C}).

We used the $Gaia$ DR3 parallax method to determine the absolute parameters of the AP Dor system.

HR and Mass-Radius ($M-R$) diagrams show the components' evolutionary state (Figure \ref{Fig5}a,b). Both the primary and secondary stars of AP Dor lie between the Zero-Age Main Sequence (ZAMS) and the Terminal-Age Main Sequence (TAMS). The position of AP Dor on the $R_{ratio}-q$ relationship provided by \cite{2022PASP..134f4201P} is also depicted in Figure \ref{Fig5}c. In addition, the $logJ_0-logM$ diagram shows the position of the system (Figure \ref{Fig5}d) and this diagram shows that AP Dor is in a contact binary systems region. According to calculations, the orbital angular momentum of AP Dor has a value of $51.847\pm0.046$. This result is based on the equation presented by \cite{2006MNRAS.373.1483E} as follows:

\begin{equation}\label{eq5}
J_0=\frac{q}{(1+q)^2} \sqrt[3] {\frac{G^2}{2\pi}M^5P}
\end{equation}

where $q$ is the mass ratio, $M$ is the total mass, $P$ is the orbital period, and $G$ is the gravitational constant.
\\
\\
\cite{2004AA...416.1097S} were analysed this system with the aid of Rucinski's simplified light curve synthesis method (\citealt{1993PASP..105.1433R}). As written in the conclusion section of this study, the method used for analysis is used for large databases of variables observed with moderate accuracy, as in the case of the $HIPPARCOS$ mission photometry (\citealt{1997AJ....113..407R}). So, they no attempt has been made to use more sophisticated light curve solution methods. Therefore, the \cite{2004AA...416.1097S} study estimated a mass ratio $q=0.1$, a fillout factor $f=1$, and an inclination $i=62.5$. According to \cite{2004AA...416.1097S}'s study, the value of the fillout factor for the AP Dor system seems unrealistic due to the difference in temperature between the components, which has not reached equilibrium.

There is a significant disparity between the findings of the \cite{2004AA...416.1097S} study and those of this investigation. The method used in the \cite{2004AA...416.1097S} study, a large number of investigated systems, and the estimation of only three main parameters show that their results are controversial.

Our results show that the short orbital period and light curve analysis of AP Dor demonstrate that it is an overcontact eclipsing binary with a fillout factor of $48.8\%$ and a mass ratio of 0.584.

%%%%%%%%%%%%%%%%%%%%%%%%%%%%%%%%%%%%%%%%%%%%%%%%%%

\section*{Acknowledgements}
This manuscript was prepared by the Binary Systems of South and North (BSN) project (\url{https://bsnp.info/}).
We have made use of data from the European Space Agency (ESA) mission Gaia (\url{http://www.cosmos.esa.int/gaia}), processed by the Gaia Data Processing and Analysis Consortium (DPAC).
This work includes data from the TESS mission observations. Funding for the TESS mission is provided by the NASA Explorer Program.
We would like to thank Filiz Kahraman Aliçavuş, and Paul D. Maley for their scientific assistance.

%%%%%%%%%%%%%%%%%%%%%%%%%%%%%%%%%%%%%%%%%%%%%%%%%%

\section*{ORCID iDs}
\noindent Atila Poro: 0000-0002-0196-9732\\
Eduardo Fernández Lajús: 0000-0002-9262-4456\\
Mohammad Madani: 0000-0003-4705-923X\\
Golshan Sabbaghian: 0000-0002-0615-4292\\
Farshid Nasrollahzadeh: 0000-0003-4444-8942\\
Faezeh Jahediparizi: 0000-0002-6813-8124\\

%%%%%%%%%%%%%%%%%%%% REFERENCES %%%%%%%%%%%%%%%%%%

%%%%%%%%%%%%%%%%%%%% APPENDIX %%%%%%%%%%%%%%%%%%

\appendix
\section{Available Minima Times}
The available minima extracted from the TESS data and literature are presented in the appendix. The first two mid-times marked with an asterisk (*) are related to the \cite{2017OEJV..179....1J} study.

\bsp	% typesetting comment

\clearpage

\setcounter{table}{0}
\begin{table*}
\caption{Available mid-eclipse times of AP Dor obtained by CCD.}
\centering
\begin{center}
\footnotesize
\begin{tabular}{c c c c c c c c c}
 \hline
 \hline
Min.($BJD_{TDB}$) & Epoch & O-C & Min.($BJD_{TDB}$) & Epoch & O-C & Min.($BJD_{TDB}$) & Epoch & O-C\\
\hline
2456966.7203(20)*	&	-2549.0	&	-0.00244	&	2458366.61655(5)	&	728.0	&	0.00201	&	2458380.92820(6)	&	761.5	&	0.00289	\\
2457603.87174(30)*	&	-1057.5	&	-0.00041	&	2458366.83170(7)	&	728.5	&	0.00357	&	2458381.14208(3)	&	762.0	&	0.00318	\\
2458354.22792(6)	&	699.0	&	0.00180	&	2458367.04222(11)	&	729.0	&	0.00049	&	2458381.35636(12)	&	762.5	&	0.00387	\\
2458354.44207(4)	&	699.5	&	0.00236	&	2458368.75246(8)	&	733.0	&	0.00198	&	2458386.05700(12)	&	773.5	&	0.00545	\\
2458354.65602(5)	&	700.0	&	0.00272	&	2458368.96785(6)	&	733.5	&	0.00378	&	2458386.26741(4)	&	774.0	&	0.00227	\\
2458354.86943(5)	&	700.5	&	0.00253	&	2458369.17978(4)	&	734.0	&	0.00212	&	2458386.48194(6)	&	774.5	&	0.00320	\\
2458355.08305(4)	&	701.0	&	0.00256	&	2458369.39508(6)	&	734.5	&	0.00382	&	2458386.69472(4)	&	775.0	&	0.00239	\\
\hline
\hline
\end{tabular}
\end{center}
\label{A1}
\end{table*}

\begin{table*}
\renewcommand\thetable{1}
\caption{Continued}
\centering
\begin{center}
\footnotesize
\begin{tabular}{c c c c c c c c c}
 \hline
 \hline
Min.($BJD_{TDB}$) & Epoch & O-C & Min.($BJD_{TDB}$) & Epoch & O-C & Min.($BJD_{TDB}$) & Epoch & O-C\\
\hline
2458399.51099(5)	&	805.0	&	0.00305	&	2458417.02517(8)	&	846.0	&	0.00256	&	2458435.39453(6)	&	889.0	&	0.00288	\\
2458399.72426(4)	&	805.5	&	0.00273	&	2458417.24010(8)	&	846.5	&	0.00390	&	2458435.60875(6)	&	889.5	&	0.00351	\\
2458399.93824(5)	&	806.0	&	0.00311	&	2458417.45234(7)	&	847.0	&	0.00255	&	2458435.82181(5)	&	890.0	&	0.00298	\\
2458400.15154(5)	&	806.5	&	0.00282	&	2458417.66712(6)	&	847.5	&	0.00373	&	2458468.50114(10)	&	966.5	&	0.00250	\\
2458400.36551(4)	&	807.0	&	0.00320	&	2458417.87966(6)	&	848.0	&	0.00268	&	2458468.71501(6)	&	967.0	&	0.00278	\\
2458400.57857(5)	&	807.5	&	0.00266	&	2458418.09429(7)	&	848.5	&	0.00372	&	2458468.92938(5)	&	967.5	&	0.00355	\\
2458400.79269(5)	&	808.0	&	0.00319	&	2458418.30720(6)	&	849.0	&	0.00303	&	2458469.14213(4)	&	968.0	&	0.00271	\\
\hline
\hline
\end{tabular}
\end{center}
\label{A2}
\end{table*}

\begin{table*}
\renewcommand\thetable{1}
\caption{Continued}
\centering
\begin{center}
\footnotesize
\begin{tabular}{c c c c c c c c c}
 \hline
 \hline
Min.($BJD_{TDB}$) & Epoch & O-C & Min.($BJD_{TDB}$) & Epoch & O-C & Min.($BJD_{TDB}$) & Epoch & O-C\\
\hline
2458481.95863(5)	&	998.0	&	0.00360	&	2458522.11349(5)	&	1092.0	&	0.00288	&	2458541.12369(7)	&	1136.5	&	0.00326	\\
2458482.17195(6)	&	998.5	&	0.00333	&	2458522.32820(7)	&	1092.5	&	0.00400	&	2458541.33771(7)	&	1137.0	&	0.00369	\\
2458482.38587(5)	&	999.0	&	0.00365	&	2458522.54076(5)	&	1093.0	&	0.00296	&	2458541.55082(6)	&	1137.5	&	0.00320	\\
2458482.59910(5)	&	999.5	&	0.00329	&	2458522.75549(6)	&	1093.5	&	0.00410	&	2458541.76582(12)	&	1138.0	&	0.00461	\\
2458482.81318(4)	&	1000.0	&	0.00378	&	2458522.96789(5)	&	1094.0	&	0.00291	&	2458571.23818(15)	&	1207.0	&	0.00107	\\
2458483.02636(6)	&	1000.5	&	0.00336	&	2458523.18264(6)	&	1094.5	&	0.00406	&	2458571.45469(7)	&	1207.5	&	0.00398	\\
2458483.24048(4)	&	1001.0	&	0.00389	&	2458523.39503(6)	&	1095.0	&	0.00286	&	2458571.66826(6)	&	1208.0	&	0.00396	\\
\hline
\hline
\end{tabular}
\end{center}
\label{A3}
\end{table*}

\begin{table*}
\renewcommand\thetable{1}
\caption{Continued}
\centering
\begin{center}
\footnotesize
\begin{tabular}{c c c c c c c c c}
 \hline
 \hline
Min.($BJD_{TDB}$) & Epoch & O-C & Min.($BJD_{TDB}$) & Epoch & O-C & Min.($BJD_{TDB}$) & Epoch & O-C\\
\hline
2458586.61954(6)	&	1243.0	&	0.00369	&	2458660.09531(4)	&	1415.0	&	0.00330	&	2458673.97905(4)	&	1447.5	&	0.00346	\\
2458586.83314(7)	&	1243.5	&	0.00370	&	2458660.31026(4)	&	1415.5	&	0.00466	&	2458674.19367(4)	&	1448.0	&	0.00449	\\
2458587.04670(6)	&	1244.0	&	0.00367	&	2458660.52243(4)	&	1416.0	&	0.00323	&	2458674.40625(5)	&	1448.5	&	0.00348	\\
2458587.26025(8)	&	1244.5	&	0.00362	&	2458660.73756(5)	&	1416.5	&	0.00477	&	2458674.62080(4)	&	1449.0	&	0.00443	\\
2458587.47406(6)	&	1245.0	&	0.00384	&	2458660.94972(4)	&	1417.0	&	0.00334	&	2458674.83346(5)	&	1449.5	&	0.00350	\\
2458587.68748(6)	&	1245.5	&	0.00367	&	2458661.16481(5)	&	1417.5	&	0.00483	&	2458675.04803(4)	&	1450.0	&	0.00448	\\
2458587.90116(5)	&	1246.0	&	0.00375	&	2458661.37690(5)	&	1418.0	&	0.00333	&	2458675.26075(5)	&	1450.5	&	0.00360	\\
\hline
\hline
\end{tabular}
\end{center}
\label{A4}
\end{table*}

\begin{table*}
\renewcommand\thetable{1}
\caption{Continued}
\centering
\begin{center}
\footnotesize
\begin{tabular}{c c c c c c c c c}
 \hline
 \hline
Min.($BJD_{TDB}$) & Epoch & O-C & Min.($BJD_{TDB}$) & Epoch & O-C & Min.($BJD_{TDB}$) & Epoch & O-C\\
\hline
2459040.93422(4)	&	2306.5	&	0.00500	&	2459054.81781(4)	&	2339.0	&	0.00501	&	2459123.16827(4)	&	2499.0	&	0.00555	\\
2459041.14861(4)	&	2307.0	&	0.00580	&	2459055.03231(5)	&	2339.5	&	0.00592	&	2459123.38292(5)	&	2499.5	&	0.00661	\\
2459041.36134(5)	&	2307.5	&	0.00493	&	2459055.24489(4)	&	2340.0	&	0.00491	&	2459123.59525(4)	&	2500.0	&	0.00534	\\
2459041.57567(5)	&	2308.0	&	0.00567	&	2459055.45951(5)	&	2340.5	&	0.00593	&	2459123.81010(5)	&	2500.5	&	0.00660	\\
2459041.78859(5)	&	2308.5	&	0.00500	&	2459055.67215(3)	&	2341.0	&	0.00498	&	2459124.02238(5)	&	2501.0	&	0.00529	\\
2459042.00264(4)	&	2309.0	&	0.00545	&	2459055.88669(5)	&	2341.5	&	0.00592	&	2459124.23744(4)	&	2501.5	&	0.00675	\\
2459042.21595(4)	&	2309.5	&	0.00517	&	2459056.09940(3)	&	2342.0	&	0.00504	&	2459124.44963(4)	&	2502.0	&	0.00535	\\
\hline
\hline
\end{tabular}
\end{center}
\label{A5}
\end{table*}

\begin{table*}
\renewcommand\thetable{1}
\caption{Continued}
\centering
\begin{center}
\footnotesize
\begin{tabular}{c c c c c c c c c}
 \hline
 \hline
Min.($BJD_{TDB}$) & Epoch & O-C & Min.($BJD_{TDB}$) & Epoch & O-C & Min.($BJD_{TDB}$) & Epoch & O-C\\
\hline
2459139.18815(4)	&	2536.5	&	0.00592	&	2459154.99472(5)	&	2573.5	&	0.00657	&	2459174.85843(4)	&	2620.0	&	0.00608	\\
2459139.40209(4)	&	2537.0	&	0.00627	&	2459155.20792(4)	&	2574.0	&	0.00618	&	2459175.07327(5)	&	2620.5	&	0.00733	\\
2459139.61534(4)	&	2537.5	&	0.00592	&	2459155.42190(4)	&	2574.5	&	0.00656	&	2459175.50047(4)	&	2621.5	&	0.00734	\\
2459139.82923(4)	&	2538.0	&	0.00622	&	2459155.63521(4)	&	2575.0	&	0.00628	&	2459175.71253(5)	&	2622.0	&	0.00581	\\
2459140.04254(5)	&	2538.5	&	0.00594	&	2459155.84911(4)	&	2575.5	&	0.00659	&	2459175.92768(5)	&	2622.5	&	0.00737	\\
2459140.25636(4)	&	2539.0	&	0.00616	&	2459156.06236(4)	&	2576.0	&	0.00624	&	2459176.13946(4)	&	2623.0	&	0.00555	\\
2459140.46972(5)	&	2539.5	&	0.00593	&	2459156.27626(5)	&	2576.5	&	0.00655	&	2459176.35481(5)	&	2623.5	&	0.00731	\\
\hline
\hline
\end{tabular}
\end{center}
\label{A6}
\end{table*}

\begin{table*}
\renewcommand\thetable{1}
\caption{Continued}
\centering
\begin{center}
\footnotesize
\begin{tabular}{c c c c c c c c c}
 \hline
 \hline
Min.($BJD_{TDB}$) & Epoch & O-C & Min.($BJD_{TDB}$) & Epoch & O-C & Min.($BJD_{TDB}$) & Epoch & O-C\\
\hline
2459190.45092(5)	&	2656.5	&	0.00625	&	2459233.59750(4)	&	2757.5	&	0.00694	&	2459250.04403(4)	&	2796.0	&	0.00677	\\
2459190.66521(4)	&	2657.0	&	0.00695	&	2459233.81063(4)	&	2758.0	&	0.00648	&	2459250.25787(4)	&	2796.5	&	0.00702	\\
2459190.87797(7)	&	2657.5	&	0.00611	&	2459234.23797(4)	&	2759.0	&	0.00663	&	2459250.47118(4)	&	2797.0	&	0.00674	\\
2459191.09246(4)	&	2658.0	&	0.00701	&	2459234.45195(5)	&	2759.5	&	0.00702	&	2459250.68524(4)	&	2797.5	&	0.00720	\\
2459191.30521(4)	&	2658.5	&	0.00617	&	2459234.66535(4)	&	2760.0	&	0.00682	&	2459250.89839(4)	&	2798.0	&	0.00676	\\
2459191.51969(4)	&	2659.0	&	0.00705	&	2459234.87910(4)	&	2760.5	&	0.00698	&	2459251.11222(4)	&	2798.5	&	0.00700	\\
2459191.73258(6)	&	2659.5	&	0.00635	&	2459235.09230(3)	&	2761.0	&	0.00659	&	2459251.32542(4)	&	2799.0	&	0.00660	\\
\hline
\hline
\end{tabular}
\end{center}
\label{A7}
\end{table*}

\begin{table*}
\renewcommand\thetable{1}
\caption{Continued}
\centering
\begin{center}
\footnotesize
\begin{tabular}{c c c c c c c c c}
 \hline
 \hline
Min.($BJD_{TDB}$) & Epoch & O-C & Min.($BJD_{TDB}$) & Epoch & O-C & Min.($BJD_{TDB}$) & Epoch & O-C\\
\hline
2459372.43460(7)	&	3082.5	&	0.00827	&	2459378.84263(7)	&	3097.5	&	0.00849	&	2459384.60917(6)	&	3111.0	&	0.00801	\\
2459372.64720(6)	&	3083.0	&	0.00727	&	2459379.05640(5)	&	3098.0	&	0.00867	&	2459384.82241(7)	&	3111.5	&	0.00765	\\
2459372.86173(6)	&	3083.5	&	0.00821	&	2459379.26974(7)	&	3098.5	&	0.00842	&	2459385.24963(8)	&	3112.5	&	0.00769	\\
2459373.07436(5)	&	3084.0	&	0.00725	&	2459379.48371(5)	&	3099.0	&	0.00879	&	2459385.46356(5)	&	3113.0	&	0.00802	\\
2459373.28886(6)	&	3084.5	&	0.00815	&	2459379.69696(7)	&	3099.5	&	0.00845	&	2459385.67708(8)	&	3113.5	&	0.00795	\\
2459373.71590(7)	&	3085.5	&	0.00801	&	2459380.12411(7)	&	3100.5	&	0.00841	&	2459386.10442(7)	&	3114.5	&	0.00810	\\
2459373.92872(5)	&	3086.0	&	0.00723	&	2459380.33809(6)	&	3101.0	&	0.00880	&	2459386.31797(5)	&	3115.0	&	0.00806	\\
\hline
\hline
\end{tabular}
\end{center}
\label{A8}
\end{table*}

%%%%%%%%%%%%%%%%%%%%%%%%%%%%%%%%%%%%%%%%%%%%%%%%%%

\label{lastpage}
\end{document}